\documentclass[prl,amsmath,superscriptaddress,twocolumn,showpacs]{revtex4-1}
\usepackage{bm}
\usepackage{amssymb}
\usepackage{colordvi}
\usepackage{graphicx}
\usepackage{color}
\usepackage{hyperref}

\newcommand{\beq}{\begin{equation}}
\newcommand{\eeq}{\end{equation}}
\newcommand{\bea}{\begin{eqnarray}}
\newcommand{\eea}{\end{eqnarray}}

\newcommand{\ome}{\omega}

\def\grad {\mbox{\boldmath$\nabla$\unboldmath}}

\begin{document}

\title{Topological Node-Lines in Mechanical Metacrystals}

\author{Zhan Xiong}\email{These authors contributed equally.}
\affiliation{College of Physics, Optoelectronics and Energy, \&
  Collaborative Innovation Center of Suzhou Nano Science and
  Technology, Soochow University, 1 Shizi Street, Suzhou 215006,
  China}
\author{Hai-Xiao Wang}\email{These authors contributed equally.}
\affiliation{College of Physics, Optoelectronics and Energy, \&
  Collaborative Innovation Center of Suzhou Nano Science and
  Technology, Soochow University, 1 Shizi Street, Suzhou 215006,
  China}
\author{Jinjie Shi}
\affiliation{College of Physics, Optoelectronics and Energy, \&
  Collaborative Innovation Center of Suzhou Nano Science and
  Technology, Soochow University, 1 Shizi Street, Suzhou 215006,
  China}
\author{Jie Luo}
\affiliation{College of Physics, Optoelectronics and Energy, \&
  Collaborative Innovation Center of Suzhou Nano Science and
  Technology, Soochow University, 1 Shizi Street, Suzhou 215006,
  China}
\author{Yun Lai}%\email{laiyun@suda.edu.cn}
\affiliation{College of Physics, Optoelectronics and Energy, \&
  Collaborative Innovation Center of Suzhou Nano Science and
  Technology, Soochow University, 1 Shizi Street, Suzhou 215006,
  China}
\author{Ming-Hui Lu}\email{luminghui@nju.edu.cn}
\affiliation{National Laboratory of Solid State Microstructures, \& Department of Materials Science and Engineering, Nanjing University, Nanjing, Jiangsu 210093, China}
\author{Jian-Hua Jiang}\email{jianhuajiang@suda.edu.cn, joejhjiang@hotmail.com}
\affiliation{College of Physics, Optoelectronics and Energy, \&
  Collaborative Innovation Center of Suzhou Nano Science and
  Technology, Soochow University, 1 Shizi Street, Suzhou 215006,
  China}

\date{\today}

\begin{abstract}
Topological acoustic and elastic waves have recently emerged as an exciting interdisciplinary field which is still mainly 
focused on low-dimensional structures and model systems. Here we demonstrate numerically an elastic-wave analogue of  
topological node-lines in three-dimensional mechanical metacrystals with ribbon- 
or drumhead-like surface states. These two-dimensional topological surface states offer unprecedented, robust subwavelength 
confinement of elastic waves. Design principles for topological mechanical metamaterials, 
from both material and symmetry aspects, are unveiled and connected to fundamental conservation laws and nonsymmorphic 
space group. Our study paves the way toward the synergy between three-dimensional 
mechanical metamaterials and topological wave dynamics.
\end{abstract}

\maketitle

{\sl Introduction}.---The discovery of topological insulators and quantized edge transport has renewed our understanding of quantum phases
of condensed matters \cite{a1,a2}. Recently, the exploration of topological physics has been extended from electronic \cite{a1,a2} and matter 
\cite{at-review} waves to classical waves such as acoustic \cite{ac1,ac2} and photonic \cite{ph-review} waves. As benefited from 
good controllability and measurability in broad frequency ranges and enriched by its vectorial nature, classical waves emerge 
as intriguing media for the study of topological phenomena \cite{ac1,ac2,ac3,ac4,ac5,ac6,ph-review,el-1,el0,gauge,el1,el2,el3,el5,el6,el7,el8,el9,el-zak}. Photonic, acoustic, and elastic topological 
edge states provide robust wave propagation which are ideal for guiding energy and information flow against noisy, imperfect environments. 

\begin{figure}
\begin{center}
\includegraphics[width=8cm]{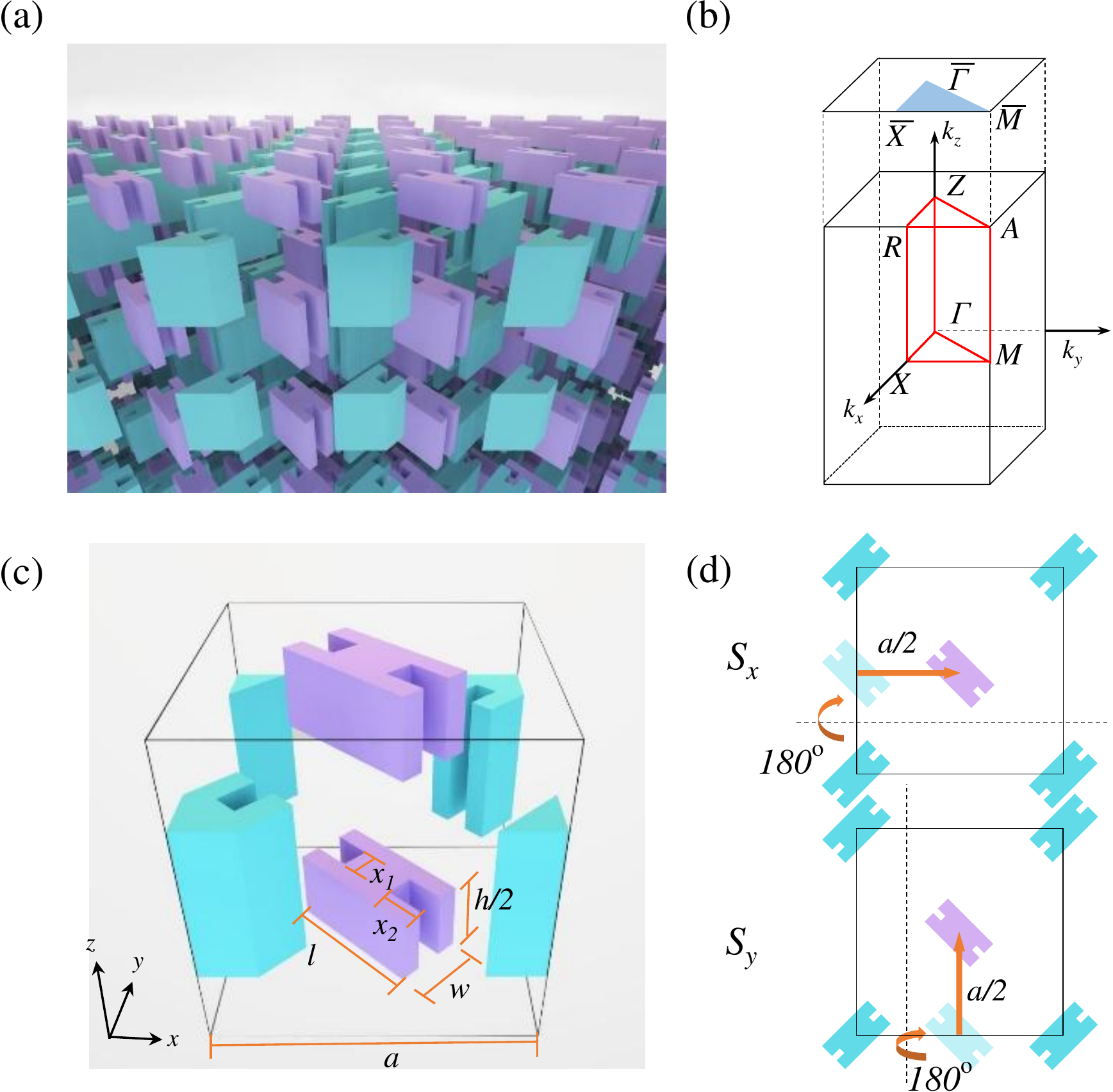}
\caption{ (Color online) (a) 3D view of the mechanical metacrystal. (b) The first BZs for the bulk and surface elastic waves. (c) The unit cell structure of the metacrystal, comprised of the background material (transparent) and the scatters (the purple and the cyan structures). (d) Illustration of the nonsymmorphic screw symmetries.}
\end{center}
\end{figure}

%In contrast to the much attention devoted to photonic and acoustic systems, there are much fewer studies on topological 
%phenomena in elastic waves \cite{el-1,el0,gauge,el1,el2,el3,el5,el6,el7,el8,el9,el-zak}, most of which are focused on low-dimensional 
%mechanical systems. Interesting phenomena such as states of self-stress \cite{el-1,el0}, synthetic gauge fields \cite{gauge}, and 
%phononic edge states \cite{el-1,el1,el2,el3,el5,el6,el7,el9,el-zak}, have been demonstrated.

To date, the rich physics of three-dimensional (3D) topological elastic waves remains unexplored 
(with only two recent exceptions \cite{3d1,3d2}). Owing to the full polarization degrees of freedom and the larger wavevector space, 
3D mechanical waves can support versatile topological states that do not have analog in low-dimensional systems \cite{3d1,3d2}. Surprisingly, 
3D mechanical metacrystals and metamaterials, despite their important roles and very broad range applications in
the cutting-edge material technologies \cite{wegner}, have not yet been considered as hosts for topological elastic waves.

In this Letter, we present numerical discovery of mechanical topological node-lines in a class of 3D mechanical metacrystals 
of tetragonal symmetry. A topological node-line is a line-degeneracy between two bands in 3D wavevector space as described by \cite{burkov}
\beq
{\cal H} =\ome_0 + \delta {\vec k}_{\perp}\cdot \hat{v} \cdot {\vec \sigma} .
\eeq
Here $\ome_0$ is the frequency of a degeneracy point on the node-line. $\delta {\vec k}$ denotes the difference wavevector with respect to 
the wavevector of the degeneracy point, where $\delta {\vec k}_\perp$ is its component perpendicular to the tangent of the node-line. 
$\hat{v}$ represents the group velocity tensor, and ${\vec \sigma}$ is the Pauli-matrix vector.

We find that the mechanical node-lines give rise to topological edge states which enable unprecedented subwavelength confinement 
of elastic waves on 2D surfaces of 3D mechanical metamaterials. The emergent edge states manifest ribbon- or drumhead-like dispersions, 
which slow down the surface wave propagation. A unique partner switching scenario is uncovered, which leads to node-lines guaranteed by the 
crystalline symmetry and fundamental conservation laws. Design principles for 3D topological mechanical metacrystals are unveiled, 
which paves the avenue toward the synergy between 3D mechanical metamaterials and topological phenomena --- an interdisciplinary field
full of opportunities for fundamental researches and applications.

{\sl Mechanical metacrystal architecture}.---Consider a tetragonal mechanical metacrystal with lattice constant $a$ along all three directions [Fig.~1(a)]. The Brillouin zones (BZs) for the bulk and surface states are given in Fig.~1(b). The metacrystal architecture is designed from both the ``scatters" and the space symmetry aspects. There are two ``scatters" of ``H" shape in each unit-cell [Fig.~1(c)], as inspired by the tuning fork. The height, length, and width of the scatters are chosen as $h=0.5a$, $l=0.5a$, and $w=0.2a$, separately. The two geometrical parameters characterizing the cut-in's of the H-shaped scatters are $x_1=0.06a$ and $x_2=0.1a$. The two scatters (labeled by different colors in Fig.~1) are of identical shape and material, but are oriented and positioned differently. The space group is the tetragonal group P42/mcm which contains the mirror operation, $\hat{M}_z: z\to -z$, and the two screw operations, $\hat{S}_x: (x,y,z)\to (\frac{a}{2}+x, \frac{a}{2}-y, \frac{a}{2}-z)$ and $\hat{S}_y: (x,y,z)\to (\frac{a}{2}-x, \frac{a}{2}+y, \frac{a}{2}-z)$. The nonsymmorphic screw operations are combinations of rotations and half-lattice translations [Fig.~1(d)]. Besides, the metacrystal also has ${\cal P}{\cal T}$ symmetry. The scatters are made of air or steel, while the background is made of an anisotropic elastic medium such as elastic metamaterials \cite{eff} or crystalline materials \cite{gypsum}. The anisotropic elasticity is important for a sizable partial band gap, as shown in the next section. The phononic band structures are calculated by solving the dynamic equations for harmonic elastic waves,
\beq
\grad \cdot \{\hat{C} [\grad {\vec \varphi} + (\grad {\vec \varphi})^T] \} = - 2 \rho \ome^2 {\vec \varphi} 
\eeq
where $\hat{C}$ is the elastic modulus tensor, ${\vec \varphi}$ is the displacement field, $\rho$ is the mass density, and $\ome$ is the angular frequency of the harmonic wave. The frequency $\ome$ scales linearly with the inverse of the lattice constant, $1/a$. For concreteness, we consider the situation with $a=1$~cm. We use the COMSOL Multiphysics software to solve the above master equation with real material properties. The elastic modulus of the background material is of the polar anisotropic form 
\beq
\hat{C} = \left( \begin{array}{ccccccccc}
     C_{11} & C_{12} & C_{13} & 0 & 0& 0\\
     C_{12} & C_{22} & C_{23} & 0 & 0& 0\\
     C_{13} & C_{23} & C_{33} & 0 & 0& 0\\
     0 & 0 & 0 & C_{44} & 0 & 0\\
    0 & 0 & 0 & 0 & C_{55} & 0\\
    0 & 0 & 0 & 0 & 0 & C_{66}
    \end{array}\right) ,
\eeq
with $C_{11}=C_{22}$%=2.85\times 10^{10}$~N/m$^2$, $C_{33}=8.58\times 10^9$~N/m$^2$
, $C_{13}=C_{23}$%=4.29\times 10^9$~N/m$^2$, $C_{44}=1.5\times 10^9$~N/m$^2$
, $C_{55}=C_{66}$%=1.0\times 10^{10}$~N/m$^2$
, and $C_{12}=C_{11}-2C_{66}$. %=8.5\times 10^9$~N/m$^2$ \cite{gypsum}. 
The three principal axes are the $x$, $y$, $z$ axes. The six indices in the above matrix are $1\equiv xx$, $2\equiv yy$, $3\equiv zz$, $4\equiv xy$, $5\equiv xz$, and $6\equiv yz$. %The mass density is $2.35\times 10^3$~kg/m$^3$ \cite{gypsum}. 
The material parameters are given in the Supplemental Materials \cite{sm}.

\begin{figure}
\begin{center}
\includegraphics[width=8cm]{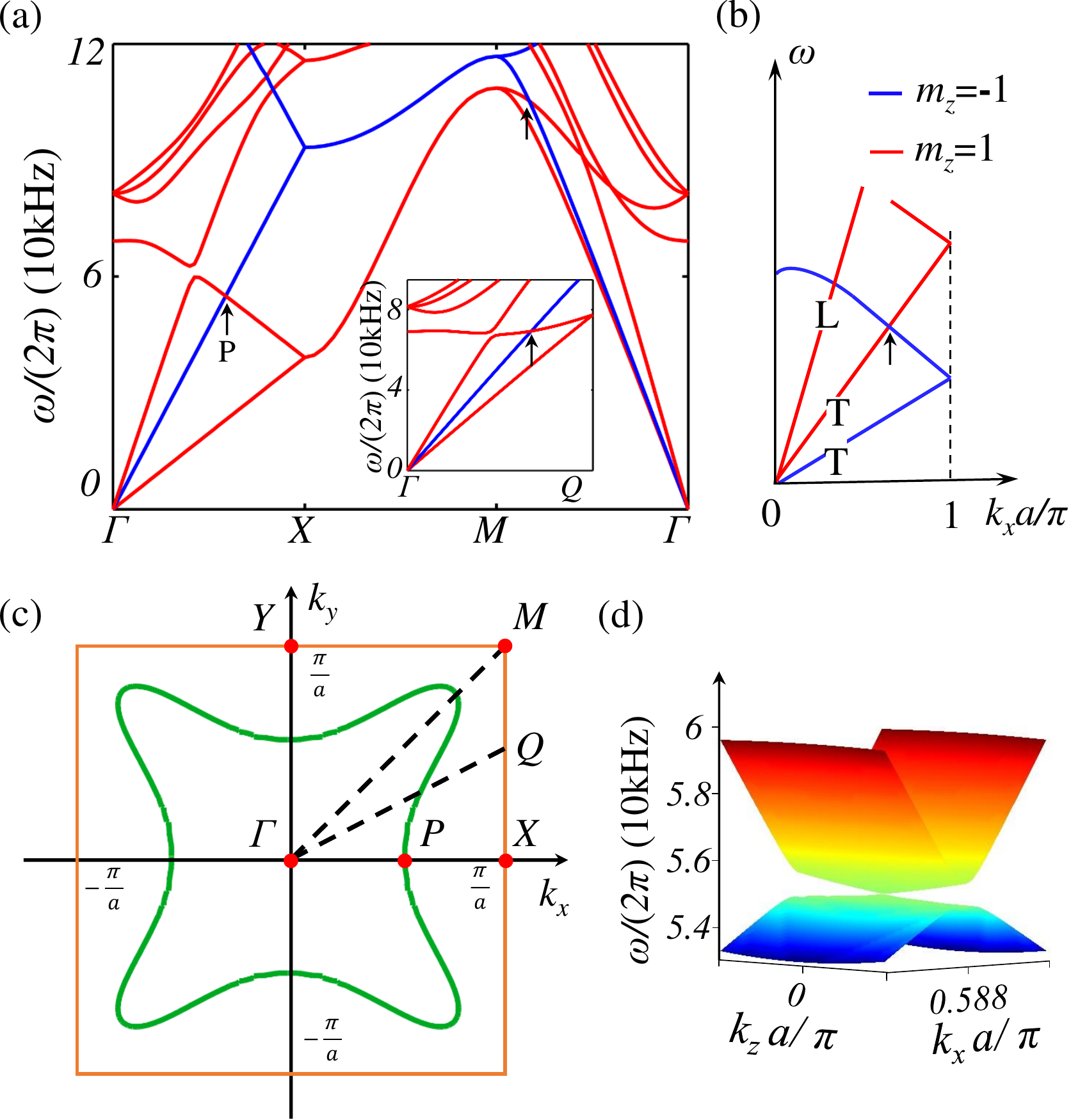}
\caption{(Color online) (a). Band structure of phononic states at the $k_z=0$ plane. Red and blue curves denote phononic states of mirror eigenvalues, $m_z=\pm 1$, respectively. Inset: band diagram on the $\Gamma$$Q$ line with $Q$ denoting the ${\vec k}=(\frac{\pi}{a},0.5\frac{\pi}{a},0)$ point. (b) Schematic depiction of band-evolution at the $k_z=0$ plane between the $\Gamma$ point and an arbitrary point at the Brillouin zone boundary for another case different from (a). In (a) and (b), the arrows indicate unavoidable degeneracies between the even and odd bands. In (a) [(b)] the transverse even branch has smaller (larger) group velocity than the odd branch. (c) The first topological node-line and the partner-switching mechanism in the $k_z=0$ plane. The dashed lines indicate the $\Gamma$$Q$ and $\Gamma$$M$ lines where band-evolution and band-degeneracy are examined in (a), beside the $\Gamma$$X$ line. The unavoidable node-line is due to the degeneracy partner switching between the $\Gamma$ point and the BZ boundary (the orange lines). (d) Dirac dispersion at the $P$ point [i.e., the intersection point between the node-line and the $k_x$ axis, as illustrated in (c)]. The metacrystal is made of gypsum \cite{gypsum} (background) and air (scatters) }
\end{center}
\end{figure}

{\sl Symmetry-induced band degeneracy}.---We now introduce a unique property of the phononic spectrum: at the BZ boundary plane $k_x=\frac{\pi}{a}$, all phononic bands are doubly degenerate. To reveal the underlying mechanism, we construct an anti-unitary operator $\hat{\Theta}_x=\hat{S}_x*\hat{{\cal T}}$ ($\hat{{\cal T}}$ is the time-reversal operator) which transforms the wavevector as $(k_x,k_y,k_z)\to (-k_x,k_y,k_z)$. Thus, $\hat{\Theta}_x$ is an invariant operator at the $k_x=\frac{\pi}{a}$ plane where it has the special property, 
\beq
\hat{\Theta}_x^2{\vec \varphi}_{n{\vec k}}=e^{-ik_xa}{\vec \varphi}_{n{\vec k}}=-{\vec \varphi}_{n{\vec k}}, 
\eeq
for an arbitrary band index $n$. According to the Kramers theorem, this property leads to double degeneracy for all phononic bands at the BZ boundary with $k_x=\frac{\pi}{a}$ [see Fig.~2(a)]. Since for tetragonal lattices the dispersion is identical if $k_x$ and $k_y$ are interchanged, the same property holds for the $k_y=\frac{\pi}{a}$ plane.

{\sl Parity evolution and mechanical node-lines}.---The topological node-line resides in the $k_z=0$ plane, where the phononic states can be labeled by the mirror eigenvalues $\hat{M}_z{\vec \varphi}_{n{\vec k}}=m_z{\vec \varphi}_{n{\vec k}}$ with $m_z=\pm 1$. We find that
\beq 
\hat{M}_z\hat{\Theta}_x{\vec \varphi}_{n{\vec k}}=e^{-ik_za}\hat{\Theta}_x\hat{M}_z{\vec \varphi}_{n{\vec k}}=m_z\hat{\Theta}_x{\vec \varphi}_{n{\vec k}},
\eeq 
for $k_z=0$. This property indicates that at the BZ boundary, the two degenerate Bloch states, ${\vec \varphi}_{n{\vec k}}$ and $\hat{\Theta}_x{\vec \varphi}_{n{\vec k}}$, have the {\em same} mirror eigenvalue [see Supplemental Materials\cite{sm} for numerical confirmation]. Fig.~2(a) indicates degeneracies between the second and the third bands of opposite mirror parity along three lines from the BZ center to boundary, $\Gamma$$X$, $\Gamma$$M$, and $\Gamma$$Q$. These degeneracies are found to form node-lines rather than Weyl or Dirac points.

To illustrate this, we examine the evolution of phononic bands. Starting from $\ome=0$, three acoustic-phonon branches evolve from the $\Gamma$ point to an arbitrary point on the $XM$ line and evolve back to form higher-frequency bands of the {\em same} $m_z$. Remarkably, there is an {\em unavoidable} crossing between the second and the third phononic bands, regardless of their group velocities [see Fig.~2(b) for a case different from Fig.~2(a)]. These arguments hold for band-evolution from the $\Gamma$ point to an {\em arbitrary} point on the BZ boundary lines at $k_z=0$ plane [e.g., see the inset of Fig.~2(a)]. %(keeping in mind that the situation for the $YM$ line is identical as for the $XM$ line for tetragonal metacrystals). 
Therefore, such unavoidable band-crossing must extend from a point to a {\em line} enclosing the $\Gamma$ point, i.e., a node-line (denoted as ``the first node-line")  [see Fig.~2(c)].
This node-line, discovered at finite frequencies, is distinct from the Weyl lines at zero frequency as found in Ref.~\cite{3d1}.

The above scenario also reveals the {\em deterministic} nature of the first node-line: it must appear due to the degeneracy-partner switch 
between the $\Gamma$ point which has degeneracy between bands of {\em opposite} $m_z$ and the BZ boundary lines which have leads degeneracy between bands of the {\em same} $m_z$. We emphasize that this scenario is unique to elastic (and electromagnetic) waves which have two branches of different mirror properties in the limit of $\ome\to 0$ and ${\vec k}\to 0$. There is no such analog in electronic or atomic systems. 

In Fig.~2(d) we present the phononic dispersion around one point [the $P$ point as labeled in Figs.~2(a) and 2(c)] on the node-line. We find that the dispersion along the tangential direction (i.e., $k_y$ direction) of the node-line is very weak, whereas along the other two directions a Dirac dispersion emerges as shown in Fig.~2(d). The Hamiltonian for the $P$ point is 
\beq
{\cal H} = \ome_0 + v_x \delta k_x\hat{\sigma}_z + v_z \delta k_z\hat{\sigma}_y,
\eeq
where $\ome_0$ is the frequency of the Dirac point (we set $\hbar\equiv 1$) and $v_x$ ($v_z$) is the group velocity along the $x$ ($z$) direction. The space-time reversal operation is manifested as $\hat{{\cal P}}\hat{{\cal T}}=\hat{\sigma}_z {\cal K}$ where $\hat{\sigma}_z$ represents the mirror operation $\hat{{\cal M}}_z$ and ${\cal K}$ is complex conjugation. The Hamiltonian satisfies Eq.~(1) and the ${\cal P}{\cal T}$ symmetry.% which forbids the appearance of $\sigma_x$ terms. 

\begin{figure}
\begin{center}
\includegraphics[width=8.7cm]{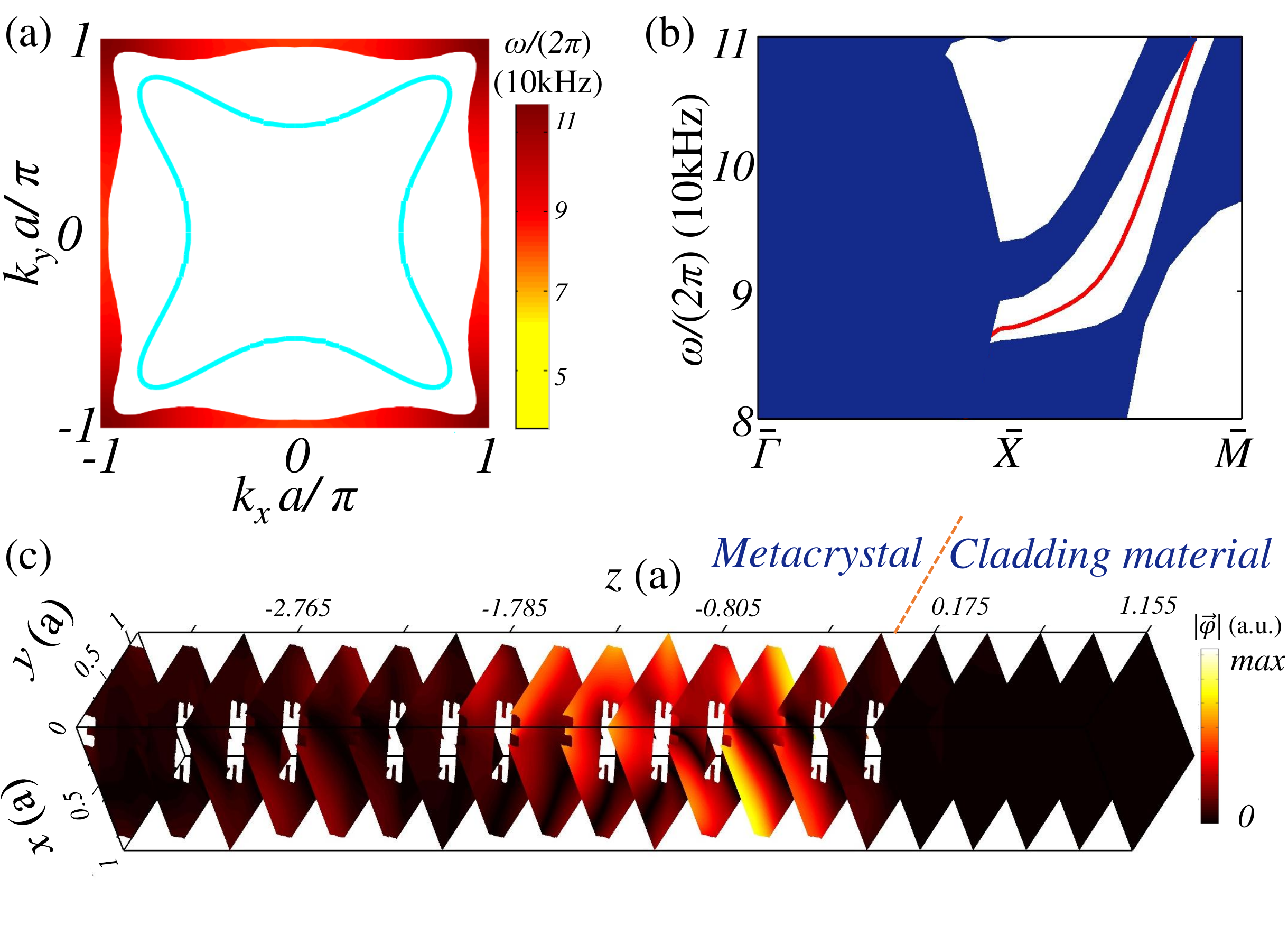}
\caption{(Color online) (a) Dispersion of the topological surface states due to the first node-line. The node-line is represented by the cyan curve. The results are obtained from a supercell calculation for (001) interface. (b) Projected bulk (blue) and surface (red) bands along the high symmetry lines $\bar{\Gamma}$$\bar{X}\bar{M}$ of the surface BZ. (c) Displacement $|{\vec \varphi}|$ profile of the topological surface wave for the ${\vec k}=\frac{\pi}{a}(1,0.6)$ point in the surface BZ with many slices along the $z$ direction. The white regions in each slice represent the scatters made of air.}
\end{center}
\end{figure}

{\sl Mechanical topological surface waves}.---The 3D phononic crystal can be reduced to the 1D Su-Schrieffer-Heeger \cite{h1} model using the dimensional reduction procedure \cite{qi}: for given $k_x$ and $k_y$ the metacrystal is equivalent to a 1D system along the $z$ direction. The ${\cal P}{\cal T}$ symmetry and the ${M}_z$ symmetry guarantee that such an effective 1D system has trivial or nontrivial Zak phase, i.e., $\theta_{Zak}(k_x,k_y)=0$ or $\pi$. It has been proven \cite{h2} 
that with these symmetries, the Zak phase for each given $(k_x,k_y)$ can be simplified into (for phononic analog, see Supplemental Materials\cite{sm})
\beq
\frac{\theta_{Zak}}{\pi}=\left\{\frac{1}{2}\sum_n \left[m_z(\frac{\pi}{a},n) - m_z(0,n)\right] \right\}{\rm mod}~ 2. 
\eeq
The first and second arguments of the $m_z$'s are the wavevector $k_z$ and the band index $n$, while the summation over $n$ includes all bands below the node-line. The node-lines are the boundaries between the regions with trivial and nontrivial Zak phases. %For instance, the Zak phase is nontrivial outside the first node-line, while inside the node-line it is trivial. 
Crossing the node-line is equivalent to a topological pumping which transfers a pair of phonon states from bulk to the two edges \cite{qi}.

The dispersion of topological edge states on the (001) surface (for calculation details, see Supplemental Materials\cite{sm}) is presented in Fig.~3(a), together with the projection of the first node-line. The topological surface states appear outside the node-line where the Zak phase is nontrivial. A finite bulk band gap is also required to observe the topological surface states. Therefore, the second and the third bands must be well-separated, which is realized by using anisotropic elastic materials. Due to the requirement of finite band-gap, topological surface states are found  in a fraction of the region outside the first node-line, forming ribbon-like spectrum with mild dispersions [see Fig.~3(a)]. The topological surface spectrum and the node-line have the $C_{4v}$ symmetry which is a projective representation of the tetragonal space group. The projected bulk bands and surface states along the high-symmetry lines are presented in Fig.~3(b). The displacement-profile of the topological surface wave for the ${\vec k}=\frac{\pi}{a}(1,0.6)$ point in the surface BZ is shown in Fig.~3(c), which indicates strong subwavelength wave confinement due to topological mechanism.

\begin{figure}
\begin{center}
\includegraphics[width=8.7cm]{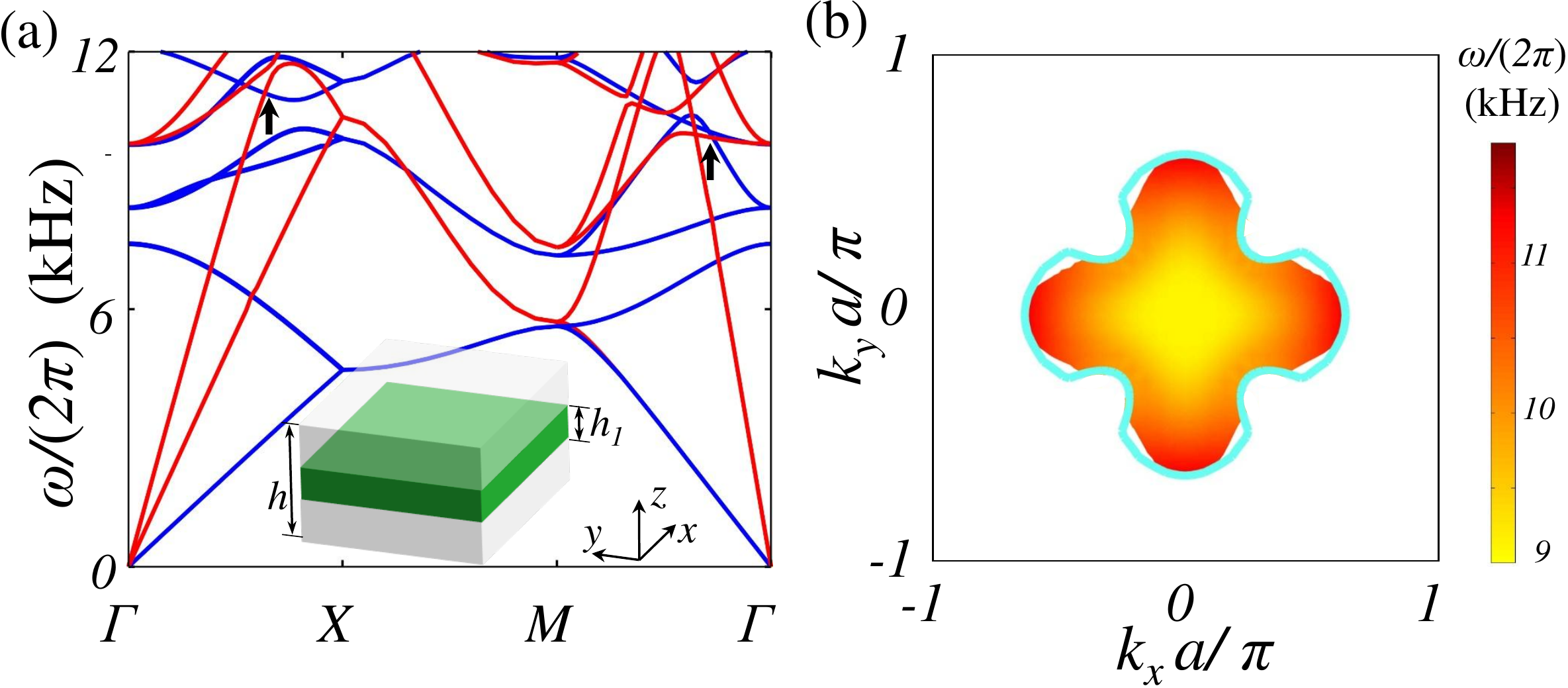}
\caption{(Color online) Phononic band structure of the $k_z=0$ plane for the mechanical metacrystal made of an elastic metamaterial (background) and steel (scatters). All geometric parameters of the metacrystal are increased to 10 times of those in Figs.~2 and 3. The elastic metamaterial is a multilayer of steel (green) and epoxy (gray) as illustrated in the inset, where $h=0.1$~cm and $h_1=0.0125$~cm. (b) Dispersion of the topological surface states (the color region) due to the node-line (depicted as the cyan curve) between the sixth and seventh bands [indicated by the arrow in (a)].}
\end{center}
\end{figure}

{\sl Node-lines from mechanical metamaterials}.---We now show that topological node-lines can also be realized by using anisotropic metamaterials as the background medium and steel as the scatters. The elastic metamaterial is made of a multilayer structure of epoxy and steel which is compatible with 3D printing technologies [see the inset of Fig.~4(a)]. The effective elastic modulus and mass density of the metamaterial are derived from the effective medium theory \cite{eff} (for details, see Supplemental Materials\cite{sm}). %, which follows the same form as in Eq. (3) for frequencies below 30kHz. 
%Such a multilayer structure is a convenient way to generate anisotropic elasticity \cite{eff}. 
We calculate the phononic band structure and topological surface states [presented in Fig.~4]. The bulk band structure shows band-evolution resembling that in Fig.~2(c). Beside the deterministic node-lines, there are accidental node-lines at higher frequencies. It turns out that the projective band gap is finite for a node-line formed by the sixth and seventh bands [indicated by the arrows in Fig.~4(a)]. The region inside this node-line has nontrivial Zak phase which leads to the drumhead-like topological surface elastic waves [see Fig.~4(b)]. The topological surface states are strongly confined around the interface and have small group velocity as indicated by their weak dispersions.
 
%{\sl Conclusion and outlook.}---Our study demonstrates crystalline-symmetry-enriched topological elastic waves in 3D mechanical 
%metacrystals which provides a versatile platform for various topological states/phenomena of elastic waves. 
%Our discovery of topological elastic waves in 3D mechanical metamaterials paves the way for the synergy
%between 3D mechanical metamaterials and topological phenomena which are desired by both physics and technology communities.

{\sl Acknowledgements.}---ZX, HXW \& JHJ acknowledge supports from the National Natural Science Foundation of China (No. 11675116). MHL acknowledges supports from the National Key R\&D Program of China (No. 2017YFA0303702) and the National Natural Science Foundation of China (No. 11625418). JS, JL \& YL thank supports from National Natural Science Foundation of China (Nos. 61671314 and 11374224).

\end{document}